\documentclass[preprint2]{aastex}

\usepackage{graphicx}
\DeclareGraphicsExtensions{.ps}  
\shorttitle{\indent \def Sausage oscillation detected by IRIS} \shortauthors{Tian et al.}

\begin{document}

\title{Global sausage oscillation of solar flare loops detected by the Interface Region Imaging Spectrograph}

\author{Hui Tian\altaffilmark{1}, Peter R. Young\altaffilmark{2,7}, Katharine K. Reeves\altaffilmark{3}, Tongjiang Wang\altaffilmark{4,7}, Patrick Antolin\altaffilmark{5}, Bin Chen\altaffilmark{6}, Jiansen He\altaffilmark{1}}
\altaffiltext{1}{School of Earth and Space Sciences, Peking University, 100871 Beijing, China; huitian@pku.edu.cn}
\altaffiltext{2}{College of Science, George Mason University, Fairfax, VA 22030, USA}
\altaffiltext{3}{Harvard-Smithsonian Center for Astrophysics, 60 Garden Street, Cambridge, MA 02138, USA}
\altaffiltext{4}{Catholic University of America, Washington, DC 20064, USA} 
\altaffiltext{5}{National Astronomical Observatory of Japan, Osawa, Mitaka, Tokyo 181-8588, Japan}
\altaffiltext{6}{Department of Physics, New Jersey Institute of Technology, Newark, NJ 07102, USA} 
\altaffiltext{7}{NASA Goddard Space Flight Center, Code 671, Greenbelt, MD 20771, USA}

\begin{abstract}
An observation from the Interface Region Imaging Spectrograph reveals coherent oscillations in the loops of an M1.6 flare on 2015 March 12. Both the intensity and Doppler shift of Fe~{\sc{xxi}}~1354.08\AA{}~show clear oscillations with a period of $\sim$25 seconds. Remarkably similar oscillations were also detected in the soft X-ray flux recorded by the Geostationary Operational Environmental Satellites (GOES). With an estimated phase speed of $\sim$2420~km~s$^{-1}$ and a derived electron density of at least 5.4$\times$10$^{10}$ cm$^{-3}$, the observed short-period oscillation is most likely the global fast sausage mode of a hot flare loop. We find a phase shift of $\sim$$\pi$/2 (1/4 period) between the Doppler shift oscillation and the intensity/GOES oscillations, which is consistent with a recent forward modeling study of the sausage mode. The observed oscillation requires a density contrast between the flare loop and coronal background of a factor $\geqslant$42. The estimated phase speed of the global mode provides an lower limit of the Alfv\'en speed outside the flare loop. We also find an increase of the oscillation period, which might be caused by the separation of the loop footpoints with time.

\end{abstract}


\section{Introduction}
Oscillations have been detected in various parts of the solar atmosphere and are often explained as signatures of magneto-hydrodynamic (MHD) waves. 
Dissipation of these waves can lead to chromospheric and coronal heating, and measurements of their characteristics can allow the diagnostics of magnetic fields and plasma properties of coronal loops \citep[e.g.,][]{Wang2016}. 

Oscillations in flares are often called quasi-periodic pulsations (QPPs). \cite{Simoes2015} found that 80\%~of the X-class flares in Cycle 24 display QPPs in the impulsive phase and that some QPPs extend into the gradual phase. \cite{Nakariakov2009} classified QPPs into short-period (sub-second) and long-period ($>$ 1 second) QPPs. Short-period QPPs are usually detected in radio and hard X-ray emission, and likely result from the interaction of plasma waves with accelerated particles. Long-period QPPs can be observed as quasi-periodic intensity fluctuation in all wavelength bands, and are likely related to MHD processes. There are two groups of possible generation mechanisms for long-period QPPs:  MHD oscillations in or near flaring loops \citep[e.g.,][]{Nakariakov2003,Nakariakov2006}, and repetitive regimes of magnetic reconnection or particle injection/acceleration \citep[e.g.,][]{Ofman2006,Barta2008,Fletcher2008}. 

When considering MHD oscillations, QPPs with periods longer than $\sim$60 seconds have been interpreted as slow magnetoacoustic mode \citep{Su2012}, global fast kink mode \citep{Kolotkov2015}, standing slow sausage mode \citep{VanDoorsselaere2011}, or fast sausage mode \citep{Srivastava2008}. While QPPs with periods of $\sim$1--60 seconds were almost exclusively interpreted as global fast sausage mode \citep{Nakariakov2003,Melnikov2005,Inglis2008,VanDoorsselaere2011,Su2012,Kolotkov2015,Chowdhury2015}, which is characterized by periodic contraction and expansion of the flux tube cross-section symmetric about the central axis \citep[e.g.,][]{Roberts1984,Cally1986,Morton2012}. The short period is mainly determined by the loop length and the large external Alfv\'en speed. 

Most observations of QPPs with periods shorter than $\sim$60 seconds have been made in radio and X-ray. Using data from the Interface Region Imaging Spectrograph \citep[IRIS,][]{DePontieu2014}, we report flare loop oscillations with a $\sim$25-second period in both the intensity and Doppler shift of the Fe~{\sc{xxi}}~1354.08\AA{}~line. Our result provides strong support to the interpretation of QPPs as global fast sausage oscillations.

\section{Observations and Results}

\begin{figure*}
\centering {\includegraphics[width=\textwidth]{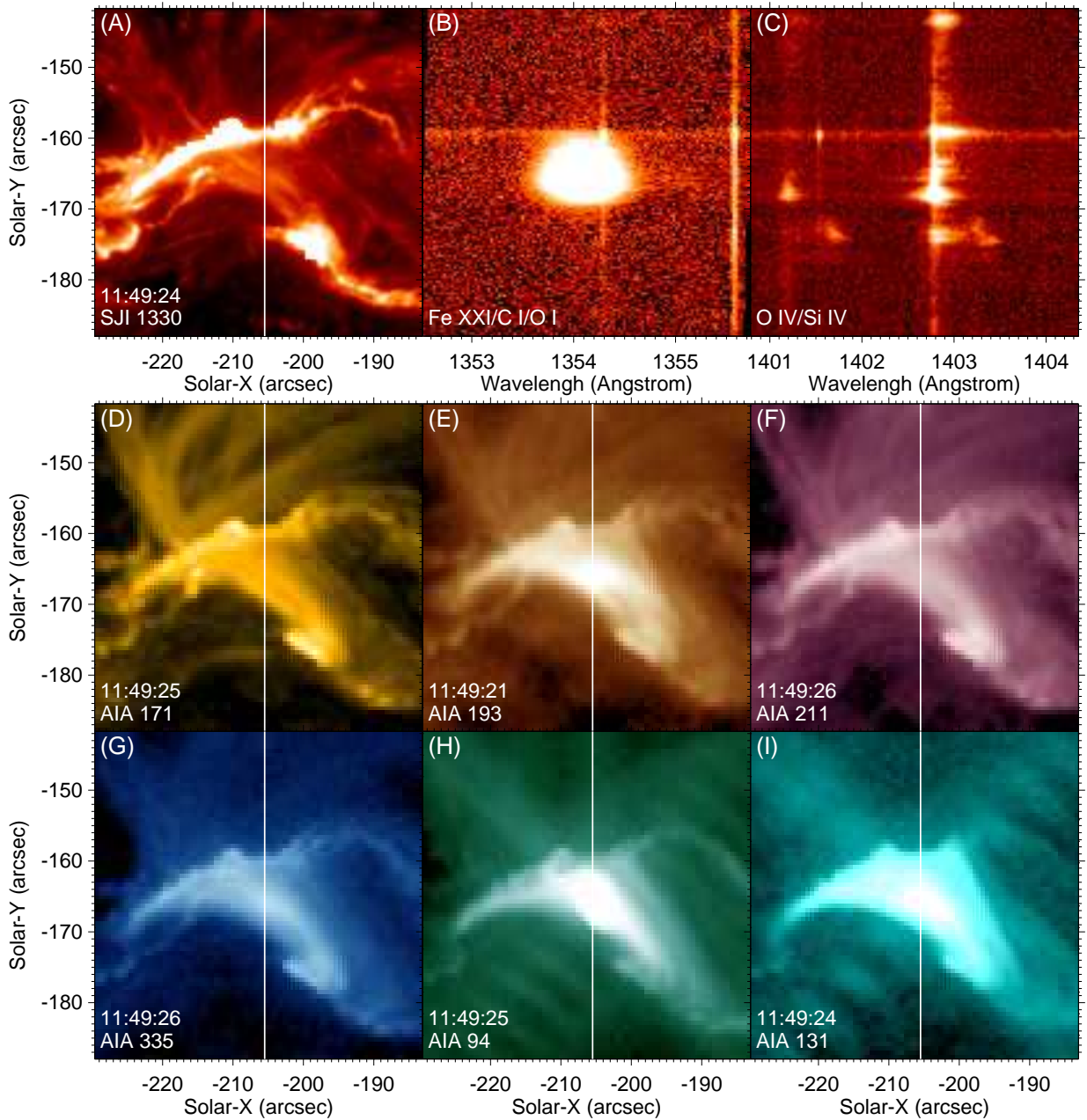}} \caption{ (A)-(C) IRIS/SJI~1330\AA{} image and detector images of the
Fe~{\sc{xxi}}~1354.08\AA{}~and Si~{\sc{iv}}~1402.77\AA{}~spectral windows at 11:49:24 UT. (D)-(I) SDO/AIA~images in the six Fe-dominated passbands taken around 11:49:24 UT. The white line in each panel indicates the slit location at the corresponding time. A movie showing the IRIS and AIA observations is available online. } \label{fig.1}
\end{figure*}

A sit-and-stare observation of IRIS was performed from 05:45 UT to 17:41 UT on 2015 March 12. The pointing coordinate was (--235$^{\prime\prime}$, --190$^{\prime\prime}$), targeting NOAA active region (AR) 12297. An M1.6 flare occurred in this AR and peaked at 11:50 UT. The data were summed by 2 both spectrally and spatially, leading to a spatial pixel size of 0.33$^{\prime\prime}$ and a spectral dispersion of $\sim$0.026 \AA{}/$\sim$0.051 \AA{} per pixel in the far/near ultraviolet wavelength bands. The cadence of the spectral observation was
$\sim$5.2 seconds, with a 4-second exposure time. Slit-jaw images (SJI) in the 1400\AA{}, 1330\AA{} and 2796\AA{} filters were taken with a cadence of $\sim$15.7 seconds for each filter. We used the level 2 data, where dark current subtraction, flat field, geometrical and orbital variation corrections have been applied. Fiducial marks on the IRIS slit were used to coalign the SJI with the spectral windows.

To examine the flare loop morphology at different temperatures we have also analyzed the data taken by the Atmospheric Imaging Assembly (AIA) onboard the Solar Dynamics Observatory (SDO). AIA images were taken at a cadence of 12 seconds in the 171\AA{}, 193\AA{}, 211\AA{}, 335\AA{}, 94\AA{} and 131\AA{} passbands and 24 seconds in the 1600\AA{} passband. The AIA images were accurately aligned to each other after applying the SolarSoft routine aia\_prep.pro, although we found that an additional manual adjustment of 2 pixels in the Solar-Y direction was necessary for the 211\AA{} and 335\AA{} channels. The IRIS images were aligned to AIA by cross-correlating AIA 1600\AA{}~(mainly ultraviolet continuum and C~{\sc{iv}}) and IRIS 1330\AA{}~(mainly ultraviolet continuum and C~{\sc{ii}}) images as they show similar spatial features. 

From Figure~\ref{fig.1} and the associated online movie we can see that the IRIS slit crossed both the flare loops and ribbons. The Fe~{\sc{xxi}}~1354.08\AA{} line, formed at $\sim$10 MK, first appears at the flare ribbon (loop footpoints) and reveals a large blue shift. This blue shift, together with the strong enhancement at the red wings of the Si~{\sc{iv}}~1402.77\AA{} and O~{\sc{iv}}~1401.16\AA{} lines (sensitive to temperatures of $\sim$0.08 MK and $\sim$0.16 MK, respectively) in the impulsive phase, indicates ongoing chromospheric evaporation. As a result of the evaporation, the Fe~{\sc{xxi}}~emission source shifts from the footpoints to the higher part of loops in the latter stage of the impulsive phase. Around the flare peak time, very strong Fe~{\sc{xxi}} emission can be identified in the loops.   

\begin{figure*}
\centering {\includegraphics[width=\textwidth]{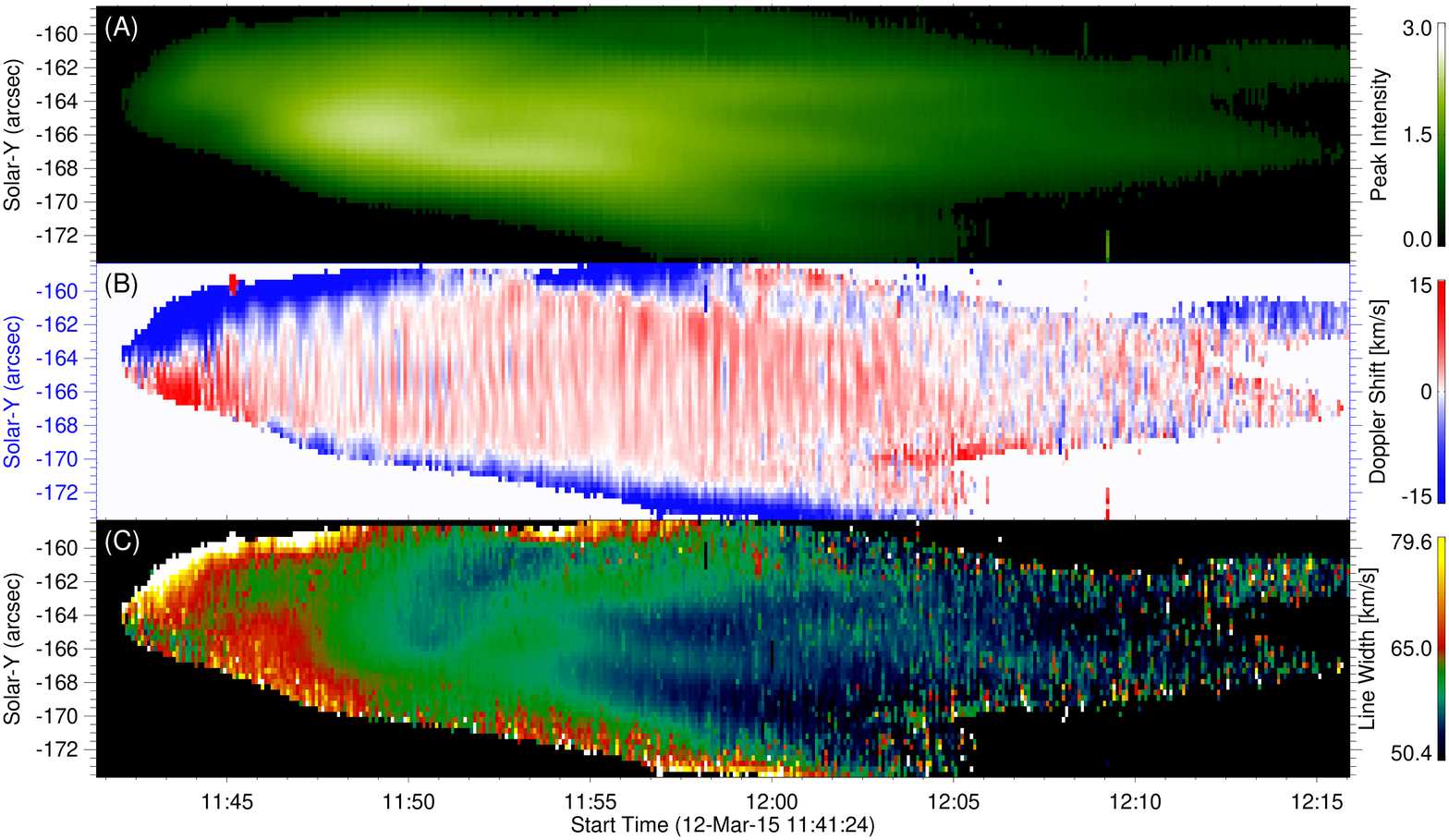}} \caption{ Fe~{\sc{xxi}}~line parameters in the flare loop. Data points with low signal to noise ratio are shown in white in the Dopplergram and black in the other images.  } \label{fig.2}
\end{figure*}

There are mainly three emission lines in the Fe~{\sc{xxi}}~spectral window: Fe~{\sc{xxi}}~1354.08\AA{}, C~{\sc{i}}~1354.29\AA{} and O~{\sc{i}}~1355.60\AA{}. For absolute wavelength calibration, we assumed that the optically thin O~{\sc{i}} line has no net Doppler shift on average. This assumption can be justified since the cold O~{\sc{i}} line has a very small intrinsic velocity. Moreover, we are mainly interested in the relative Doppler shift fluctuation, which is not affected by the uncertainty of wavelength calibration. We then applied a double-component Gaussian fit to the line profiles of Fe~{\sc{xxi}} and C~{\sc{i}} at the locations of the flare loops (from Solar-Y=--174$^{\prime\prime}$ to --158$^{\prime\prime}$), and the Fe~{\sc{xxi}}~line parameters are shown in Figure~\ref{fig.2}. There we see an obvious oscillation pattern in the Doppler shift. The oscillation appears to be largely coherent over a wide range on the slit, suggesting that different loops oscillate as a whole. Note that the large blue shifts at the northern and southern boundaries are mainly results of chromospheric evaporation, which is not analyzed in this paper. 

\begin{figure*}
\centering {\includegraphics[width=\textwidth]{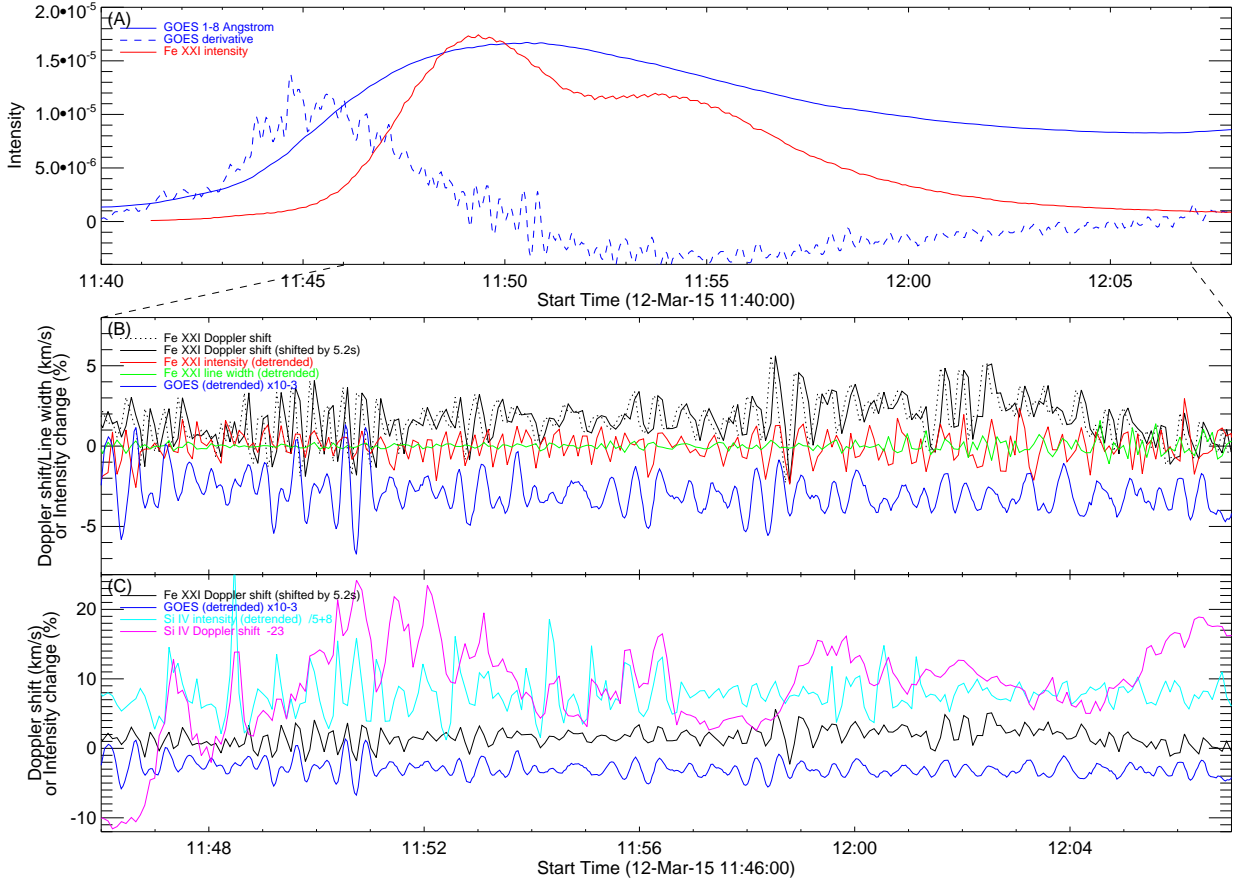}} \caption{ (A) Temporal evolution of Fe~{\sc{xxi}}~intensity, GOES 1--8\AA{}~flux and its time derivative. (B) GOES flux (detrended), Fe~{\sc{xxi}}~intensity (detrended), Doppler shift and line width (detrended). (C) GOES flux, Fe~{\sc{xxi}}~Doppler shift, Si~{\sc{iv}}~intensity (detrended) and Doppler shift. The GOES flux and Si~{\sc{iv}}~line parameters are shifted and/or rescaled for the purpose of illustration. Note that the Fe~{\sc{xxi}}~and Si~{\sc{iv}}~line parameters were derived at the loop top and ribbon, respectively. }
\label{fig.3}
\end{figure*}

We then averaged the line profiles over the loop top (from Solar-Y=--167.5$^{\prime\prime}$ to --163.8$^{\prime\prime}$) and applied the same double Gaussian fit to the averaged line profiles. Figure~\ref{fig.3} shows the time series of Fe~{\sc{xxi}} line parameters. As a comparison we also present the time series of soft X-ray flux observed with GOES and Si~{\sc{iv}}~line parameters derived through a single Gaussian fit to the line profiles averaged at the ribbon (from Solar-Y=--159.7$^{\prime\prime}$ to --158.0$^{\prime\prime}$). We also removed the trend of the intensities and line width by subtracting a smoothed (over 35 seconds) version of the time series from the original. For intensities the resultant time series was also normalized to the original time series. This process will not affect our conclusion since the dominant periods and phases, the key parameters inferred from IRIS observations for mode identification (see below), are not changed after de-trending.

Figure~\ref{fig.3}(B) reveals obvious oscillations with a period of $\sim$25 seconds in not only the Fe~{\sc{xxi}} Doppler shift, but also Fe~{\sc{xxi}} intensity and GOES flux. The intensity oscillations appear to lag the Doppler shift oscillation by one IRIS time step (5.2 s). If we shift the time series of the Doppler shift by 5.2 s, we see correlated changes of the three parameters. Since GOES records soft X-ray flux integrated over the entire Sun and the IRIS slit covers different loops, the in-phase oscillation of Fe~{\sc{xxi}} intensity and GOES flux suggests that soft X-ray emission from the flare loops dominates the GOES flux and that different loops behave as one loop. We noticed that the line width does not show significant oscillation. From Figure~\ref{fig.3}(C) we see no obvious correlation between oscillations at the loop top and ribbon. However, there is a trend of larger red shift being associated with larger intensity for Si~{\sc{iv}}.

\begin{figure}
\centering {\includegraphics[width=0.47\textwidth]{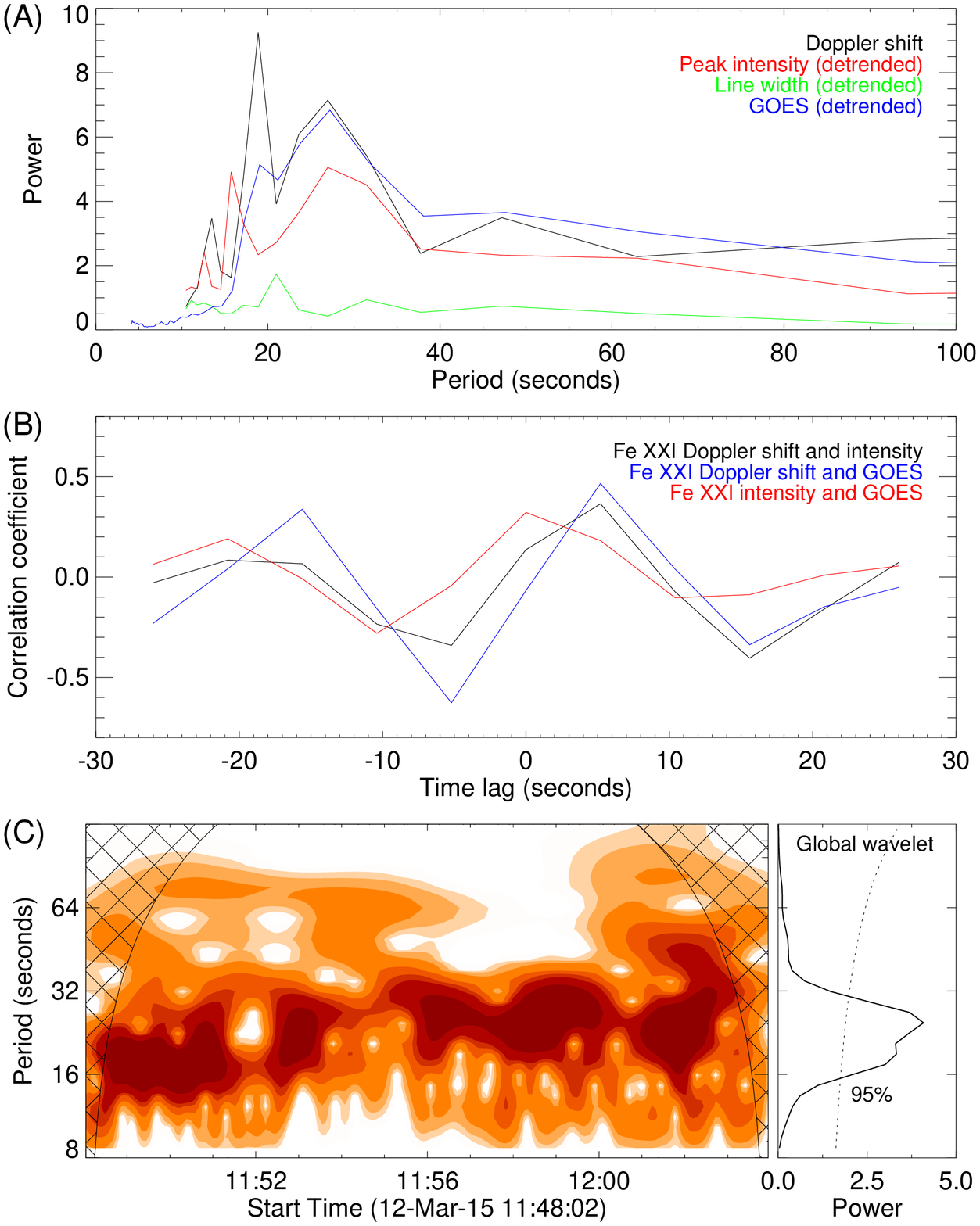}} \caption{ (A) Fourier power spectra of Fe~{\sc{xxi}}~line parameters and GOES flux. (B) Correlation coefficient between two parameters as a function of time lag. (C) Wavelet spectrum (left) and global wavelet (right) for the Fe~{\sc{xxi}}~Doppler shift. Dark color represents strong power. } \label{fig.4}
\end{figure}

We have performed Fourier analysis, wavelet analysis and cross-correlation analysis for the Fe~{\sc{xxi}}~line parameters and GOES flux in the time range of 11:48--12:04 UT (Figure~\ref{fig.4}). The Fourier power spectra reveal mainly two dominant periods, $\sim$19 seconds and $\sim$27 seconds. From the wavelet spectrum, we see that shorter periods dominate at the beginning and longer periods dominate after 11:52 UT, leading to two peaks in the global wavelet spectrum. The cross-correlation analysis confirms the phase relation we described above: the GOES flux and Fe~{\sc{xxi}}~intensity oscillations are in phase; the Fe~{\sc{xxi}}~Doppler shift leads the intensity oscillation by $\sim$5.2 seconds.  

\begin{figure*}
\centering {\includegraphics[width=0.9\textwidth]{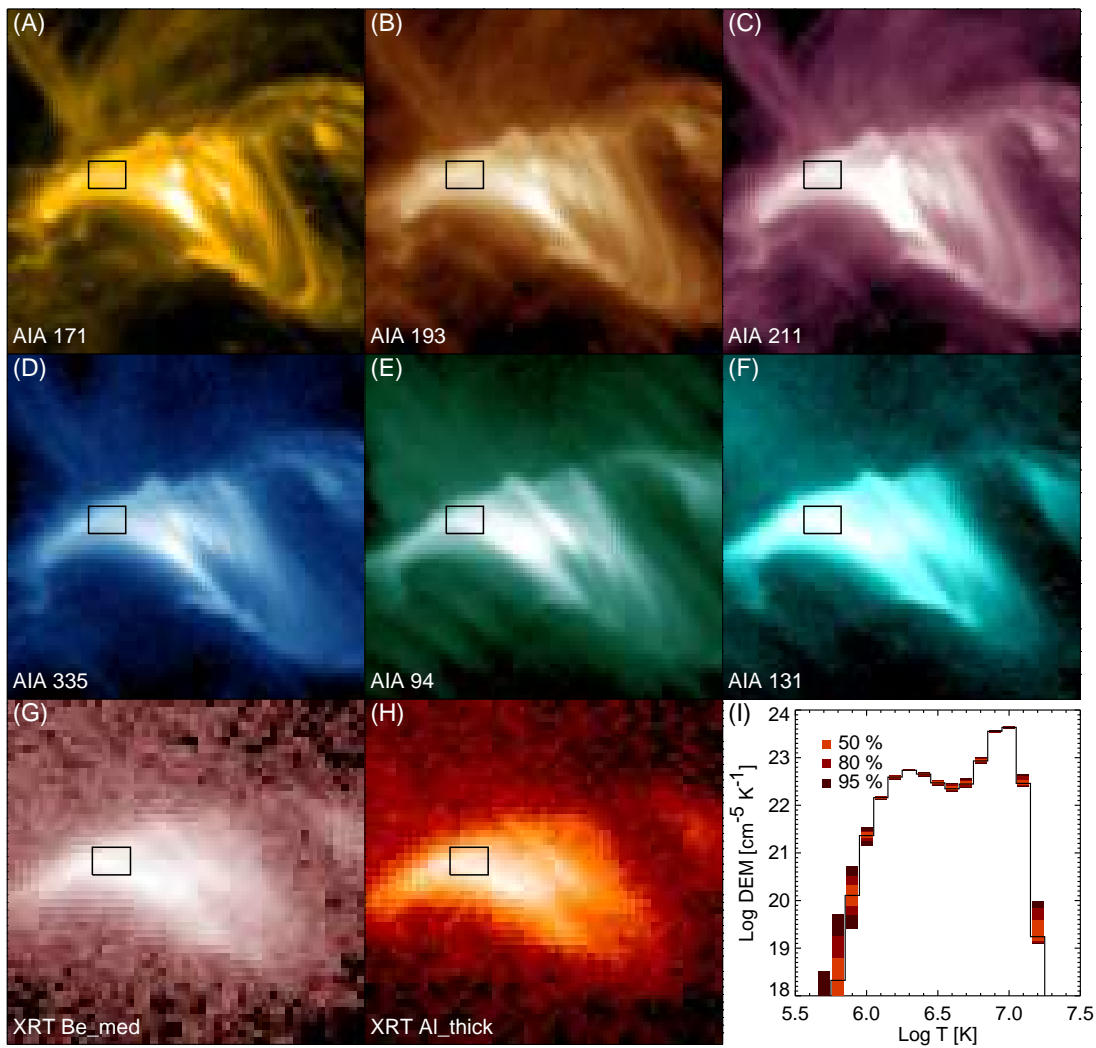}} \caption{ (A)-(H) AIA and XRT images taken around 12:02 UT. (I) The solid line shows the DEM of the flare loop, which is calculated by averaging the AIA and XRT fluxes within the rectangular region marked on the images. Boxes with different colors on the DEM plot encompass 95\%, 80\% and 50\% of the Monte Carlo solutions, respectively.} \label{fig.5}
\end{figure*}

Using AIA images taken in the six Fe-dominated passbands and images taken with the X-ray Telescope \citep[XRT,][]{Golub2007} onboard HINODE, we have performed differential emission measure (DEM) analysis for the flare loops. We chose to use images taken around 12:02 UT since XRT images in more than one filters (Be\_med and Al\_thick filters) are available around this time. Using the AIA and XRT fluxes averaged over the rectangular region marked in Figure~\ref{fig.5}(A)-(H), we have calculated the DEM using the xrt\_dem\_iterative2.pro routine in SolarSoft (SSW). This algorithm works by adjusting a series of spline knots (corresponding to different passbands) in the DEM solution until the predicted intensities are close to the observed ones \citep{Weber2004}. Errors on the DEM curve are determined by performing 100 Monte Carlo simulations, where each Monte Carlo solution is a DEM calculated using the measured intensities varied by a normally distributed random error. The Gaussian width of the normal distribution equals the uncertainty of the observed flux, which is obtained from the SSW routine aia\_bp\_estimate\_error.pro for AIA and calculated as 5\% of the observed flux for XRT. This method was originally designed for XRT data but can also be applied to the AIA data \citep{Cheng2012,Reeves2015}. The DEM shown in Figure~\ref{fig.5}(I) reveals two peaks at log ({\it T}/K)$\approx$6.3 and log ({\it T}/K)$\approx$7.0 corresponding to the average temperatures of the background corona and flare loops, respectively.

\section{Discussion}
Our observation of the Fe~{\sc{xxi}}~intensity and Doppler shift oscillations is among one of the few ultraviolet spectroscopic observations of QPPs \citep[e.g.,][]{Wang2002,Mariska2006}. Oscillations with periods of 3--6 minutes were recently identified from the transition region lines of IRIS at flare ribbons and suggested to result from some repetitive form of magnetic reconnection \citep{LiD2015,LiZhang2015,Brannon2015}. The QPPs reported here were inferred from the hot Fe~{\sc{xxi}}~line in flare loops and they have periods of only $\sim$25 seconds. 

QPPs observed in the impulsive phase of flares were often suggested to be produced by intermittent reconnection or quasi-periodic injection/acceleration of nonthermal particles \citep[e.g.,][]{Asai2001,Fleishman2008,Dolla2012}, although these processes might be modulated or triggered by MHD waves \citep[e.g., global kink mode,][]{Foullon2005}. Our QPPs are more likely to be MHD waves/oscillations, since these QPPs were mostly observed after the peak time of the GOES derivative (an approximation of the Hard X-ray) and present in the decay phase.    

The fact that different loops oscillate with the same period and almost in phase suggests that these loops oscillate as a whole and that the oscillations are most likely standing waves \citep{Melnikov2005,Inglis2008,Yu2013}. The lack of many fine structures in the hot emission from the spatially resolved loop top (Figure~\ref{fig.1} and Figure~\ref{fig.2}(A)) suggests that the hot flare plasma has a large intrinsic scale length over which the physical properties are similar, implying that the hot emission at locations of the apparent loops visible in the AIA images may actually come from a single fat magnetic loop \citep{Wang2007}. The in-phase oscillations of the spatially resolved Fe~{\sc{xxi}} intensity and the GOES flux integrated over the whole Sun further suggest the dominance of a global mode, for which the phase speed $C_{p}$ can be calculated as

\begin{equation}
\emph{$C_{p}=2L/P$}\label{equation1},
\end{equation}

where $L$ and $P$ are the loop length and oscillation period, respectively. From the AIA images, the distance $d$ between two loop footpoints was found to be $\sim$27$^{\prime\prime}$. This leads to a loop length of $\sim$42$^{\prime\prime}$ if we assume a semi-circular shape for the loops. Using the observed period of $\sim$25 seconds, $C_{p}$ was found to be $\sim$2420~km~s$^{-1}$. This speed is much higher than the sound speed $C_{s}$ at a temperature of 10 MK ($\sim$525~km~s$^{-1}$), so the QPPs are not standing slow waves. 







The QPPs are probably not consistent with a kink mode interpretation. Azimuthal flows generated by kink waves can lead to intensity enhancement. In the linear regime this is produced by column depth variation from the quadrupolar terms in the wave solution \citep{Yuan2016}, while in the nonlinear regime stronger enhancement is produced mainly by the dynamic instabilities (leading to vortices which vary the temperature and column depth) generated by the shear with the azimuthal flows \citep{Antolin2013}. However, in both regimes (and for most viewing angles) the intensity has double periodicity with respect to the Doppler shift, not consistent with our observation. Intensity variation with the same period as the Doppler shift could be produced by the variation of column depth from the geometry change of the oscillating loop \citep{Verwichte2009} or from the loop periodically crossing the slit \citep{Tian2012} in kink oscillations, and a $\pi$/2 phase shift between intensity and Doppler shift may be expected. However, in the former case phase mixing effects following the coupling of resonant absorption and dynamic instabilities would rapidly increase the phase shift to $\pi$ \citep{Antolin2015b,Okamoto2015}. In the latter scenario the slit has to be aligned along the loop, which is not true in our observation. 


Most likely, the observed QPPs are global fast sausage oscillations. The phase speed of this mode is smaller than and close to the external Alfv\'en speed $C_{Ae}$ \citep{Nakariakov2003}, which is often larger than the Alfv\'en speed in the dense flare loops (on the order of 500~km~s$^{-1}$). This interpretation is also supported by a recent forward modeling of the sausage mode \citep{Antolin2013}. In the case of a non-zero angle between the line of sight and the loop, this model predicts a $\pi$/2 phase shift between intensity and Doppler shift, as well as reduced intensity and line width fluctuations. These are all consistent with our IRIS observation. The sausage mode is often thought to have no Doppler shift oscillations \citep[e.g.,][]{Kitagawa2010}, which is not true according to this model. Actually, transition region Doppler shift oscillations with a 26-s period have been previously interpreted as fast body global sausage mode \citep{Jess2008}. The model also predicts that the period of the line width oscillation should be half of the intensity/velocity oscillation period. This can not be confirmed since the cadence of our IRIS observation is 5.2 seconds, which is not high enough to unambiguously detect oscillations with a period of $\sim$12 seconds.   

We do not see obvious damping of the QPPs during the period when the Fe~{\sc{xxi}} emission is well above the noise level. This is consistent with the theoretical prediction that the sausage oscillation is not leaky and thus showing no obvious damping when the cutoff condition, $k>k_c$ ($k$ is the wave number, equivalent to Eq.~\ref{equation2} for the fundamental mode), is satisfied \citep{VasheghaniFarahani2014,Gruszecki2012}.  

It is known that the trapped global sausage mode can only exist in dense and thick loops due to the propagation cutoff at lower wave numbers \citep{Nakariakov2003,Aschwanden2004}. \cite{Nakariakov2003} gives the following necessary condition for the existence of the global sausage mode: 

\begin{equation}
\emph{$\frac{\rho_{0}}{\rho_{e}}>(\frac{L}{0.65a})^{2}$}\label{equation2},
\end{equation}

where $a$, $\rho_{0}$ and $\rho_{e}$ are the loop cross-section diameter, internal density and external density, respectively. As mentioned above, the apparent loops visible in the AIA images are likely strands of a single fat hot loop. A recent theoretical investigation by \cite{Chen2015} also demonstrated that effects due to fine structuring can be ignored when performing coronal seismology for fast sausage oscillations. From the AIA images and the extension of Doppler shift oscillation in the slit direction, $a$ can be estimated as $\sim$10$^{\prime\prime}$. So the density contrast has to be larger than 42.  

From the DEM presented in Figure~\ref{fig.5}(I), we can estimate the loop density according to the following relation:

\begin{equation}
\emph{$\int{\it DEM}(T) dT = fN_{e}^{2}a$}\label{equation3},
\end{equation}

By assuming a filling factor ($f$) of unity and integrating the DEM curve around the high-temperature peak (log ({\it T}/K)=6.7--7.3), we found that the loop density {\it N}$_{e}$ has a lower limit of 5.4$\times$10$^{10}$ cm$^{-3}$. The density contrast is 54 if we take the typical external coronal density 10$^9$ cm$^{-3}$, satisfying the necessary condition for the existence of trapped global sausage mode.



Two dominant periods in a single oscillation were often interpreted as the fundamental mode and the second harmonic \citep{Melnikov2005,Srivastava2008,Chowdhury2015}. The period ratio between the two was often found to deviate from 2, which may result from longitudinal density stratification \citep{Andries2005}, loop expansion \citep{Verth2008} or siphon flows \citep{Li2013}. However, from the wavelet spectrum in Figure~\ref{fig.4}(C), we see that shorter periods dominate at the beginning and longer periods dominate later. Thus, we think that the shorter period of $\sim$19 seconds might not be the harmonic. Instead, the increasing dominant period might be caused by the separation of the loop footpoints with time (assuming constant $C_{p}$, see Eq.~\ref{equation1}).

Intermittent reconnection should result in similar quasi-periodic behavior of chromospheric evaporation. The lack of correspondence between the Si~{\sc{iv}} oscillations at the ribbon and Fe~{\sc{xxi}} loop oscillations suggests that intermittent reconnection is not the driver of the QPPs. Also the intermittent reconnection scenario can not explain the phase relation between the intensity and Doppler shift. In fact, the Si~{\sc{iv}} oscillations were mostly observed after the peak time of GOES derivative (11:45 UT) and thus are likely related to the episodic downward moving cold materials resulting from intermittent cooling \citep[blob-like structures or showers in coronal rain,][]{Antolin2015a}. The fact that larger red shift tends to be associated with larger intensity is also consistent with this scenario. 

\section{Summary}

Using IRIS observations, we have identified intensity and Doppler shift oscillations of Fe~{\sc{xxi}}~1354.08\AA{}~with a period of $\sim$25 seconds in the loops of an M1.6 flare. A similar oscillation has also been detected by GOES. The detected QPPs are most likely global fast sausage oscillations. Based on this interpretation, we found that the density contrast and external Alfv\'en speed are larger than 42 and larger than $\sim$2420~km~s$^{-1}$, respectively.

\begin{acknowledgements}
IRIS is a NASA small explorer mission developed and operated by LMSAL with mission operations executed at NASA Ames Research center and major contributions to downlink communications funded by ESA and the Norwegian Space Centre. Hinode is a Japanese mission developed and launched by ISAS/JAXA, with NAOJ as domestic partner and NASA and STFC (UK) as international partners. It is operated by these agencies in co-operation with ESA and NSC (Norway). This work is supported by the Recruitment Program of Global Experts of China, NSFC under grant 41574166, contract 8100002705 from LMSAL to SAO, NASA Cooperative Agreement NNG11PL10A to CUA, NASA grants NNX11AB61G, NNX13AE06G, NNX15AF48G and NNX15AJ93G.
\end{acknowledgements}

\end{document}